\documentclass[3p,11pt,twocolumn]{elsarticle}

\usepackage[utf8]{inputenc}

\usepackage{graphicx}
\usepackage{epstopdf}
\usepackage{lineno,hyperref}
\usepackage{upgreek}
\usepackage{units}
\usepackage{multirow}
\usepackage{amssymb}	
\usepackage{amsmath}	
\usepackage{textcomp}
\usepackage{pdfpages}

\usepackage{caption}
\usepackage{subcaption}

\journal{Nuclear Instruments and Methods A}

\begin{document}

\begin{frontmatter}

\title{Mapping the in-plane electric field inside irradiated diodes}
\ead{APoley@cern.ch}

\author[add1]{L. Poley}

\author[add2]{A.J. Blue}
\author[add2]{C. Buttar}
\author[add3]{V. Cindro}
\author[add4]{C. Darroch}
\author[add5]{V. Fadeyev}
\author[add6]{J. Fernandez-Tejero}
\author[add6]{C. Fleta}
\author[add5]{C. Helling}
\author[add1]{C. Labitan}
\author[add3]{I. Mandi\'{c}}
\author[add1]{S.N. Santpur}
\author[add7]{D. Sperlich}
\author[add6]{M. Ull\'{a}n}
\author[add8]{Y. Unno}

\address[add1]{Lawrence Berkeley National Laboratory, Berkeley, CA 94720, USA}
\address[add2]{SUPA School of Physics and Astronomy, University of Glasgow,
Glasgow G12 8QQ, United Kingdom}
\address[add3]{Experimental Particle Physics Department, Jožef Stefan Institute, SI-1000 Ljubljana, Slovenia}
\address[add4]{Dublin Institute of Technology, D08 X622 Dublin, Ireland}
\address[add5]{Santa Cruz Institute for Particle Physics, University of California Santa Cruz, Santa Cruz, CA 95064, USA}
\address[add6]{Instituto de Microelectrónica de Barcelona, IMB-CNM (CSIC), Campus UAB, Bellaterra, Barcelona, Spain}
\address[add7]{Physikalisches Institut, Albert-Ludwigs-Universität Freiburg, 79104 Freiburg-im-Breisgau, Germany}
\address[add8]{Institute of Particle and Nuclear Study, High Energy Accelerator Research Organization (KEK), 1-1 Oho, Tsukuba-shi, Ibaraki-ken 305-0801, Japan}


\begin{abstract}
A significant aspect of the Phase-II Upgrade of the ATLAS detector is the replacement of the current Inner Detector with the ATLAS Inner Tracker (ITk). The ATLAS ITk is an all-silicon detector consisting of a pixel tracker and a strip tracker. Sensors for the ITk strip tracker have been developed to withstand the high radiation environment in the ATLAS detector after the High Luminosity Upgrade of the Large Hadron Collider at CERN, which will significantly increase the rate of particle collisions and resulting particle tracks. During their operation in the ATLAS detector, sensors for the ITk strip tracker are expected to accumulate fluences up to $\unit[1.6\cdot10^{15}]{\text{n}_{\text{eq}}/\text{cm}^2}$ (including a safety factor of 1.5), which will significantly affect their performance. One characteristic of interest for highly irradiated sensors is the shape and homogeneity of the electric field inside its active area. 
For the results presented here, diodes with edge structures similar to full size ATLAS sensors were irradiated up to fluences comparable to those in the ATLAS ITk strip tracker and their electric fields mapped using a micro-focused X-ray beam (beam diameter \unit[$2\times3$]{$\upmu$m$^2$}). This study shows the extension and shape of the electric field inside highly irradiated diodes over a range of applied bias voltages. Additionally, measurements of the outline of the depleted sensor areas allow a comparison of the measured leakage current for different fluences with expectations for the corresponding active areas.
\end{abstract}


\end{frontmatter}


\section{Introduction}

Accompanying the High-Luminosity Upgrade of the Large Hadron Collider~\cite{HLLHC}, the ATLAS detector~\cite{ATLAS} will be upgraded accordingly. As part of the ATLAS Phase-II Upgrade~\cite{LOI}, the current Inner Detector of ATLAS will be replaced with the ATLAS Inner Tracker, consisting of silicon pixel tracker and strip trackers~\cite{TDR}.

During their operation in the ATLAS detector, sensors for the ITk strip tracker are expected to accumulate fluences up to $\unit[1.6\cdot10^{15}]{\text{n}_{\text{eq}}/\text{cm}^2}$ (including a safety factor of 1.5), which will significantly affect their performance as semiconductors.
Extensive irradiation tests have been performed using both full-size sensors and test structures. Monitoring diodes in particular are used to compare the leakage current after irradiation for estimates of the area factor, i.e. leakage current per unit area, and total leakage current of full size sensors.

The measurements presented here were conducted as a follow-up for previous studies using the same method for un-irradiated diodes of the same type and geometry used here~\cite{Diodes}. Repeating the same measurement with irradiated diodes allowed both to test the applicability of the method for highly irradiated silicon structures and to study the evolution of the active diode area with increasing fluence.

\section{Experimental setup}

Electron-hole pairs created within the depleted area of a sensor lead to an increase in the sensor current, but recombine without a current increase in the undepleted area of a sensor. The lateral extension of the depleted area can therefore be mapped using the measured diode current.

Three diodes were studied in this measurement (see figures~\ref{fig:map1a}, \ref{fig:map2a} and ~\ref{fig:map3a}), which were designed to have edge regions similar to full size sensors: n-doped strip implants in a p-doped sensor bulk were surrounded by n-doped bias and guard ring implants, while the sensor backside and edge ring were p-doped. HPK diodes used here can be assumed to have an active thickness of \unit[303~\cite{CKlein}-310~\cite{Nobu}]{$\upmu$m} and a bulk resistivity of \unit[3]{k$\Omega\cdot$cm}, IFX diodes have a thickness of \unit[300]{$\upmu$m} and a bulk resistivity of \unit[3.5]{k$\Omega\cdot$cm}. For a detailed description of the edge regions of all ATLAS17 (\cite{Miguel},~\cite{Nobu}) design diodes, see~\cite{Diodes}.

At the time of the measurements, investigations into the full depletion voltage of the used diodes after different levels of irradiation were still in progress. It was therefore decided to conduct all diode measurements presented here with an applied bias voltage of \unit[-500]{V}. Since IFX MD2 and HPK MD2 diodes were designed with similar edge structures as full ATLAS ITk strip sensors, which are foreseen to be operated at \unit[-500]{V}, this working point was chosen to provide information about the development of the depleted sensor at the sensor edge at realistic operating conditions. Information about depletion voltages from later measurements~\cite{Miguel},~\cite{Hara} is summarised in table~\ref{tab:pars}.
\begin{table}
\footnotesize
\begin{center}
\begin{tabular}{l|ccc}
 & & Current at & Induced \\
Fluence, & V$_{fD}$, & \unit[-20]{$^{\circ}$C} & current \\
$[\text{n}_{\text{eq}}/\text{cm}^2]$& $[$V$]$ & $[$A$]$ & $[$nA$]$\\
\hline
$1\cdot10^{14}$ & 300~\cite{Miguel}-400~\cite{Hara}, & $5.0\cdot10^{-7}$ & 17-18\\
$5\cdot10^{14}$ & 300~\cite{Miguel}-500~\cite{Hara} & $1.5\cdot10^{-6}$ & 14-15\\
$1\cdot10^{15}$ & $>600$~\cite{Hara} & $2.9\cdot10^{-7}$ & 12-13\\
$3\cdot10^{15}$ & $>800$~\cite{Hara} & $5.2\cdot10^{-6}$ & 9
\end{tabular}
\end{center}
\caption{Parameters of the diodes under investigation: full depleton voltages V$_{fD}$, total diode leakage current (measured for all three diodes irradiated to the same fluence together for technical reasons) and average photo current induced by X-ray beam for an individual diode.}
\label{tab:pars}
\end{table}

The test setup was flushed with nitrogen to prevent condensation on the diodes. During measurements, diodes were cooled down to \unit[-20]{$^{\circ}$C} with measured variations of unit[$\pm0.1$]{$^{\circ}$C} using a peltier-element cooling jig inside a light-sealed cold box. Precision stages allowed to move diodes under investigation with respect to an X-ray beam pointed normal to the diode's top surface in order to obtain 2D maps. A Keithley 2410 high voltage power supply was used to both bias diodes under investigation and measure the corresponding current.

\section{Performed measurements}

\begin{figure}[htp]
\begin{subfigure}{0.38\textwidth}
\centering
\includegraphics[width=\linewidth]{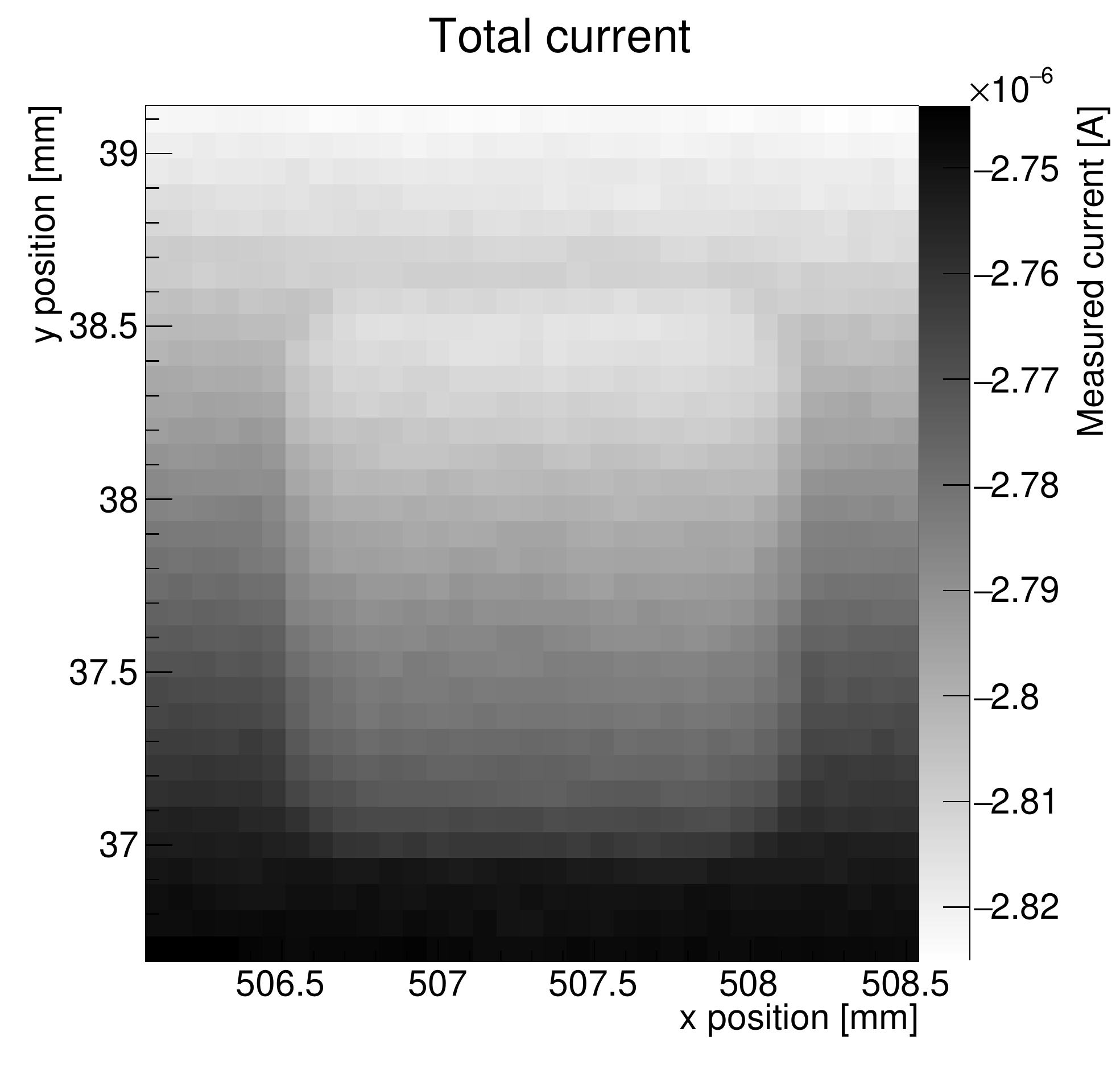}
\caption{Total diode current read out per stage position}
\label{fig:test1}
\end{subfigure}
\begin{subfigure}{0.38\textwidth}
\centering
\includegraphics[width=\linewidth]{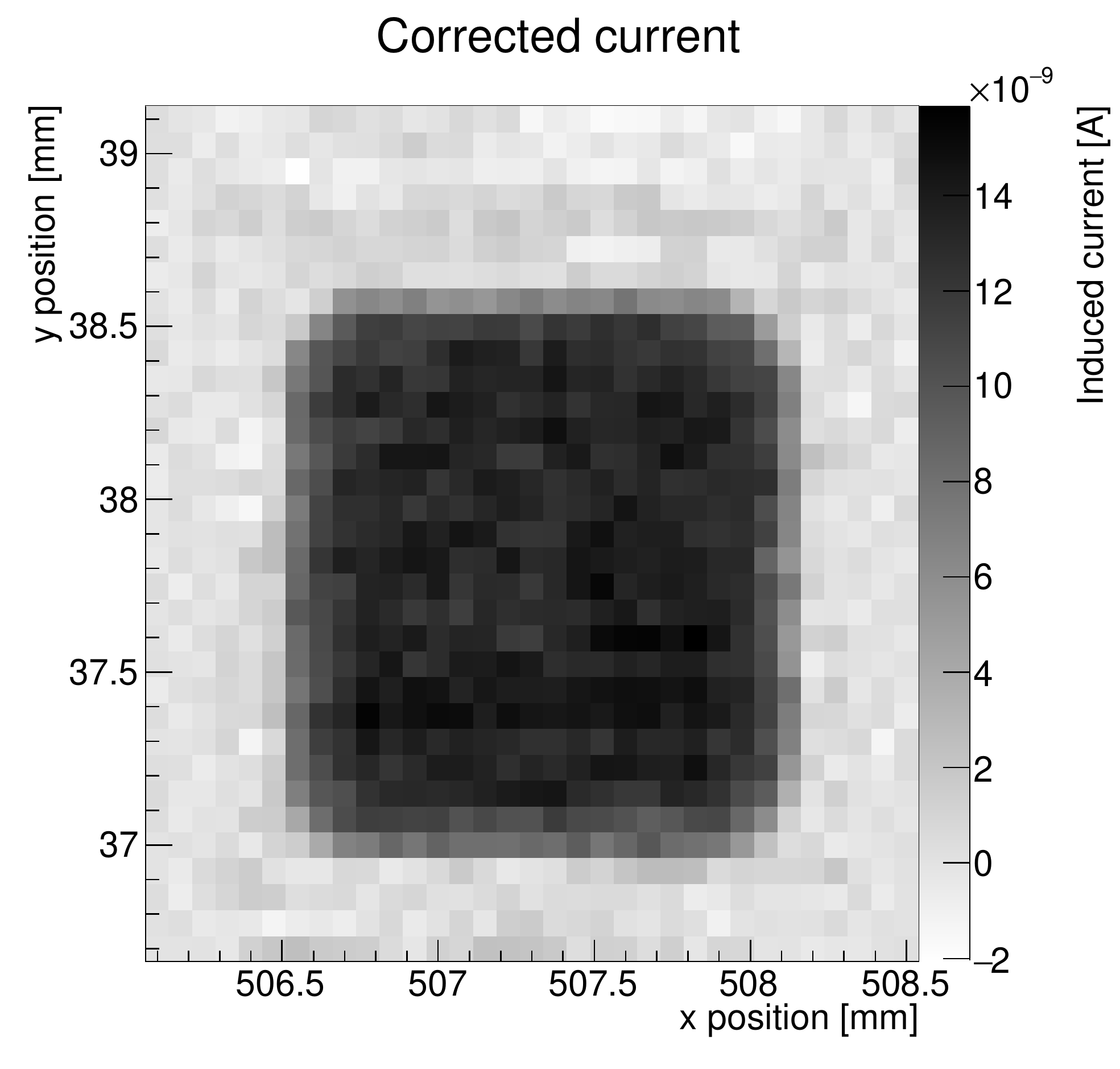}
\caption{Diode current map after subtracting diode leakage current line by line}
\label{fig:test2}
\end{subfigure}
\begin{subfigure}{0.38\textwidth}
\centering
\includegraphics[width=\linewidth]{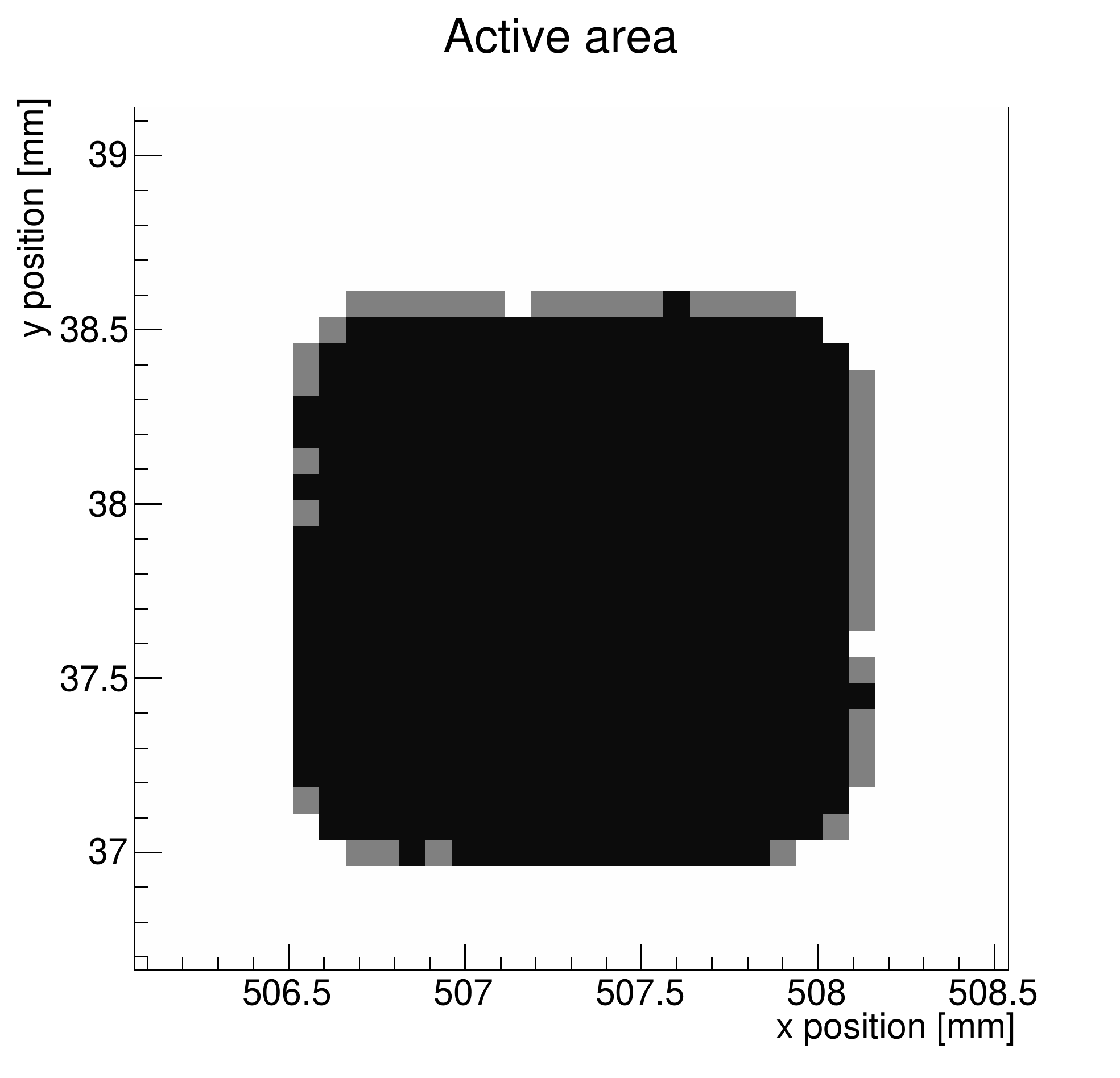}
\caption{Active area of diode calculated from bins in diode plateau: bins above the calculated threshold plus $\upsigma$ were counted towards the active area (black), bins at the threshold, but within $\upsigma$ were counted towards the active area uncertainty (grey bins, see figure~\ref{fig:test3}).}
\label{fig:test4}
\end{subfigure}
\end{figure}

Measurements were conducted using an X-ray beam focused to a beam size of $\unit[2\times3]{\upmu\text{m}^2}$ with a monochromatic beam energy of \unit[15]{keV} (beamline B16 at the Diamond Light Source). Within a diode of \unit[300]{$\upmu$m} thickness, each \unit[15]{keV} photon has a \unit[51]{\%} chance to interact with a silicon atom. The interaction produces one \unit[15]{keV} electron, which travels up to \unit[20]{$\upmu$m} within silicon, which determines the limit of the achievable position resolution.

\begin{figure}[htp]
\centering
\includegraphics[width=\linewidth]{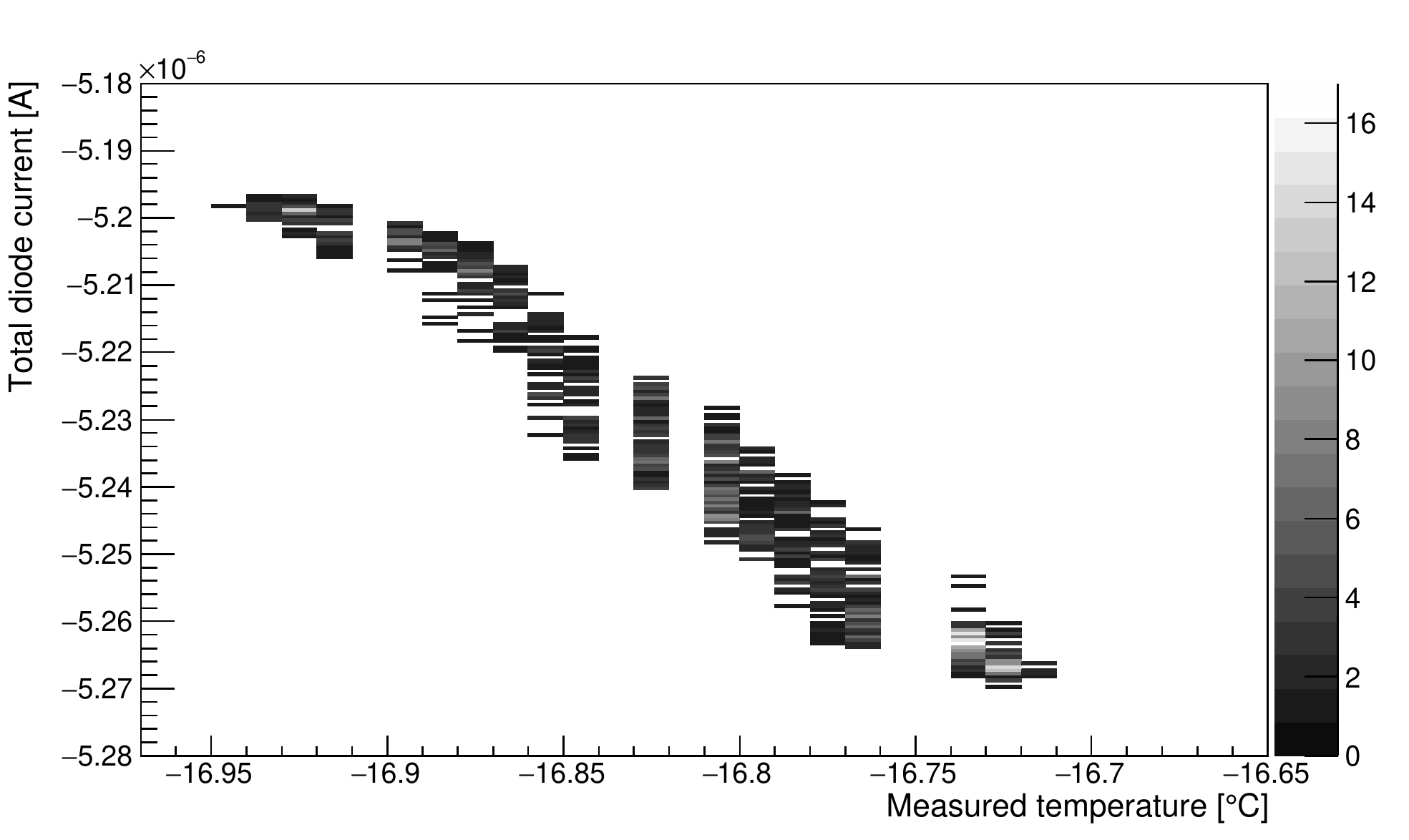}
\caption{Correlation between measured temperature and diode leakage current for a saple irradiated up to {$\unit[3\cdot10^{15}]{\text{n}_{\text{eq}}/\text{cm}^2}$}: while temperature changes of about {\unit[0.25]{$^{\circ}$C}} change the overall current by {\unit[$\pm40$]{nA}}, temperature-independent changes are significantly smaller (while including additional, beam-induced currents).}
\label{fig:correl}
\end{figure}

Diodes were scanned in a grid with a step size of $\unit[75\times75]{\upmu\text{m}^2}$ to provide good spatial resolution while minimising the beam time required per diode measurement. During measurements, the total diode current was read out. For irradiated diodes, the measured current was dominated by the sensor leakage current (see figure~\ref{fig:test1}), which varied with the sensor temperature. At a sensor temperature of about \unit[-20]{$^{\circ}$C}, actual fluctuations in the sensor leakage current, i.e. fluctuations independent of temperature changes, were small compared to the X-ray beam induced current. Figure~\ref{fig:correl}

In order to map the depleted diode area, the measured diode leakage current was subtracted from the total measured current. Scans were performed in horizontal scan lines, taken within about \unit[3]{min} per line, which can therefore be assumed to show only small fluctuations in diode temperature and overall leakage current. Therefore, a linear fit was performed for each scan line, where the un-depleted edge regions of the diode were used to calculate the total diode current per position. By subtracting the diode leakage current per bin, the underlying depleted area inside the diode was obtained (see figure~\ref{fig:test2} for an image of IFX MD2, irradiated up to $\unit[1\cdot10^{15}]{\text{n}_{\text{eq}}/\text{cm}^2}$). The resulting association of bins with the active or inactive diode area allowed to map this area directly (see figure~\ref{fig:test4}).

\begin{figure}[htp]
\centering
\includegraphics[width=\linewidth]{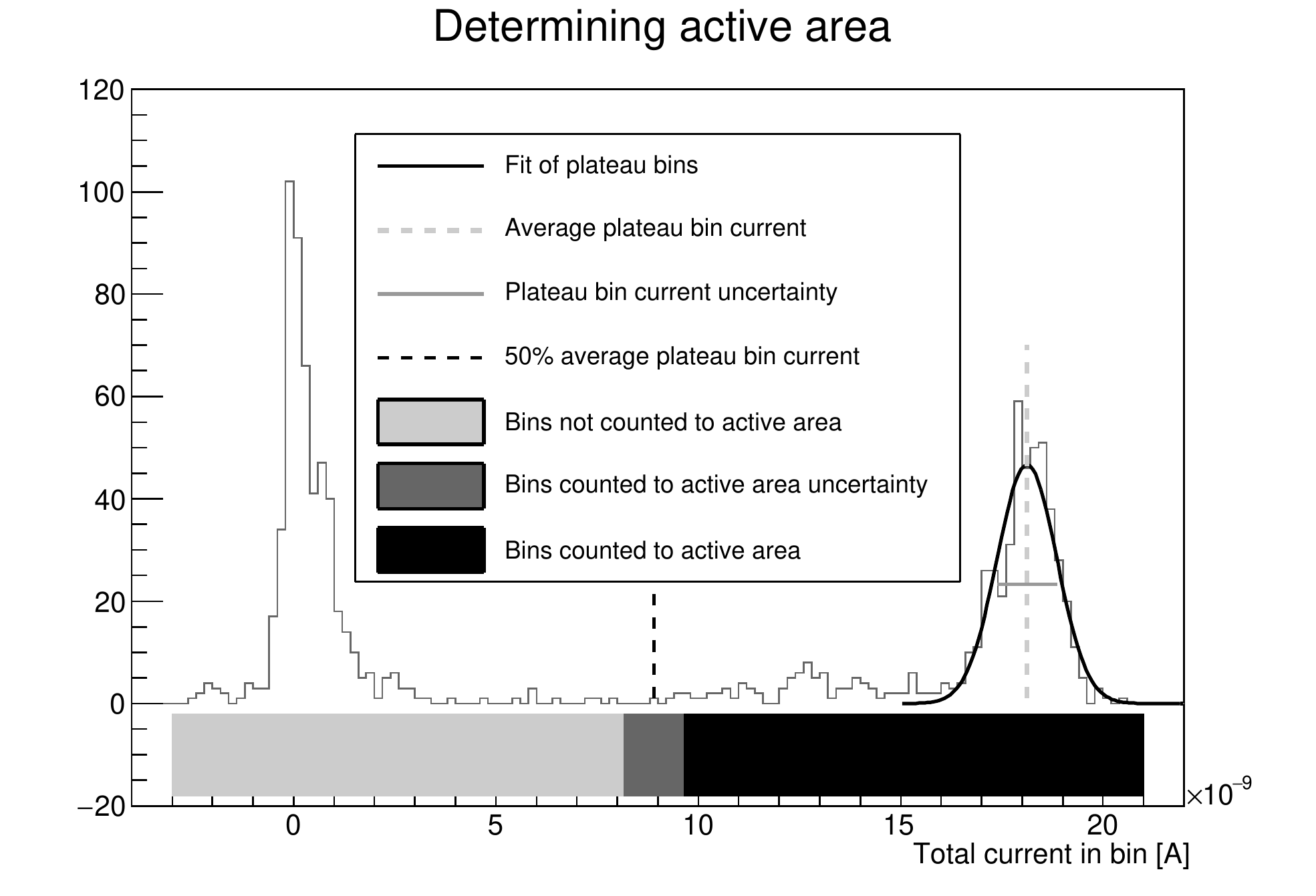}
\caption{Determining the active area of an irradiated diode from currents measured per bin: the average current measured within the active diode area was measured. Bins were counted to the active diode area if their measured current was higher than half the average plateau current.}
\label{fig:test3}
\end{figure}

The active sensor area was calculated from the resulting corrected current map: bins in the main pedestal (within fluctuations) were counted towards the active area, bins within a window within $\upsigma$ between background and pedestal were counted towards the uncertainty of the active area (see figure~\ref{fig:test3}).

\section{Results}

\begin{figure}[htp]
\begin{subfigure}{.23\textwidth}
  \centering
  \includegraphics[width=\linewidth]{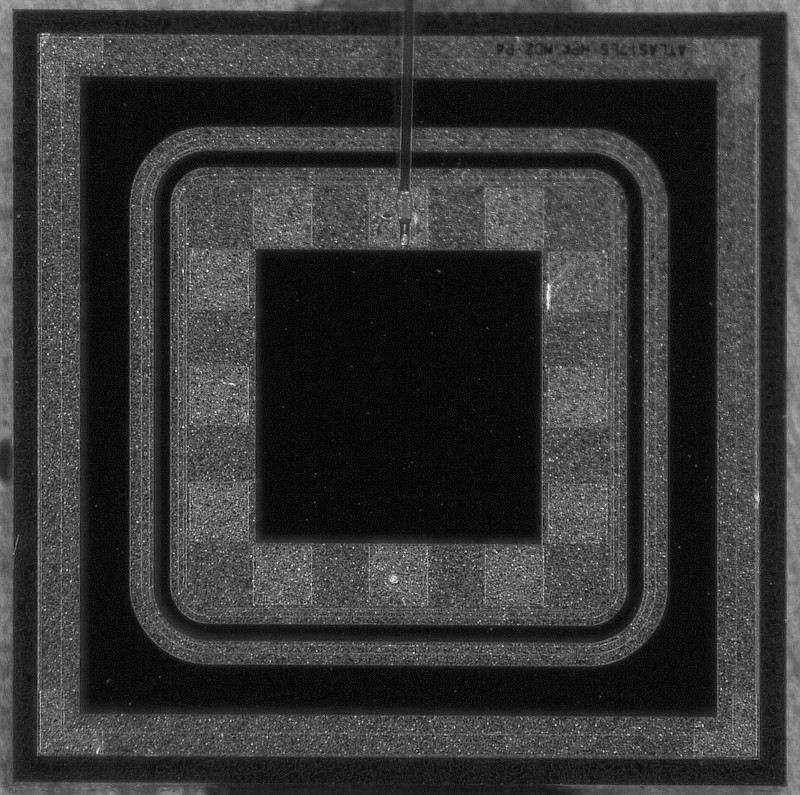}
  \caption{HPK MD2 (picture)}
  \label{fig:map1a}
\end{subfigure}
\begin{subfigure}{.23\textwidth}
  \centering
  \includegraphics[width=\linewidth]{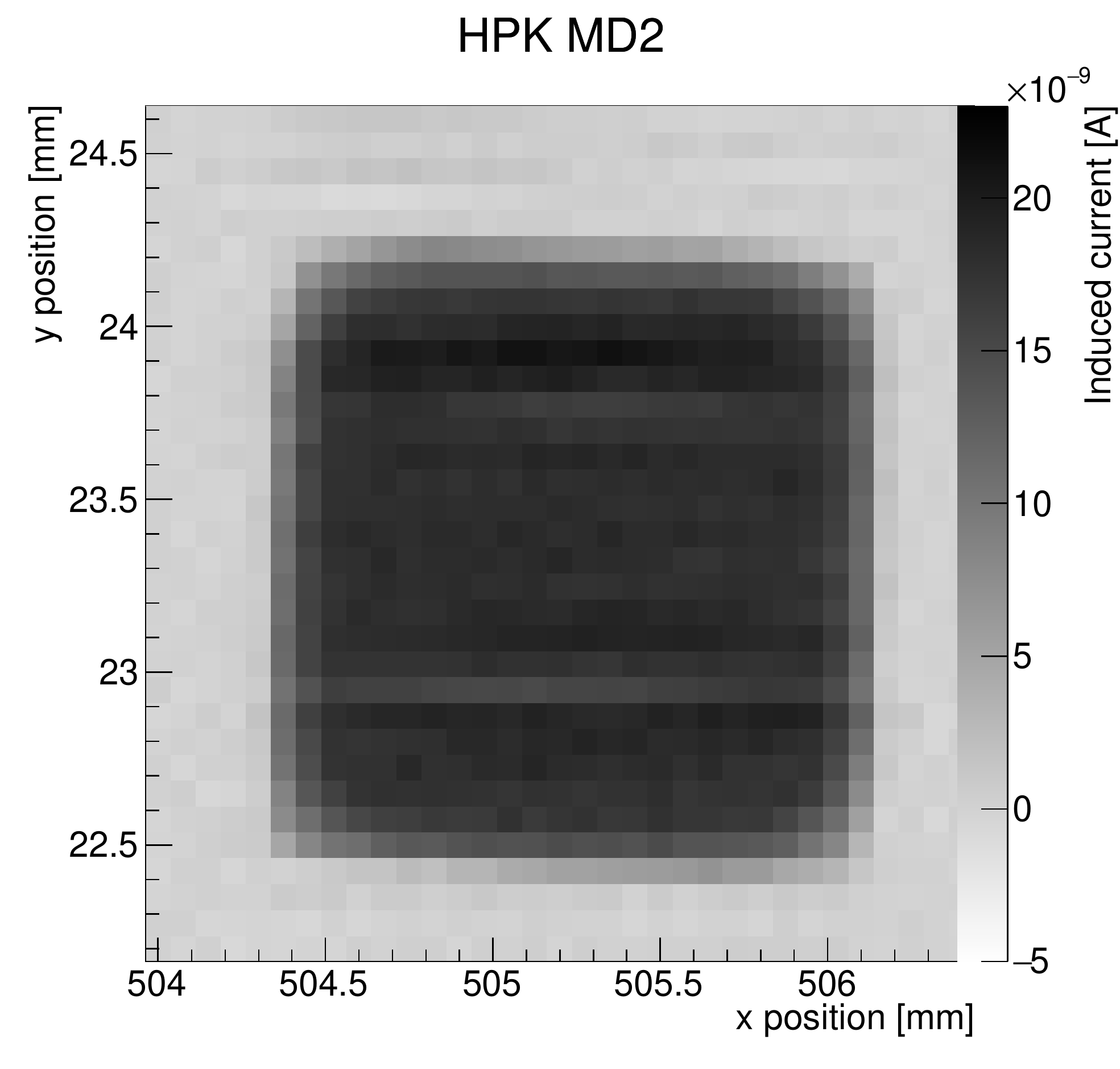}
  \caption{HPK MD2 (map)}
  \label{fig:map1b}
\end{subfigure}
\begin{subfigure}{.23\textwidth}
  \centering
  \includegraphics[width=\linewidth]{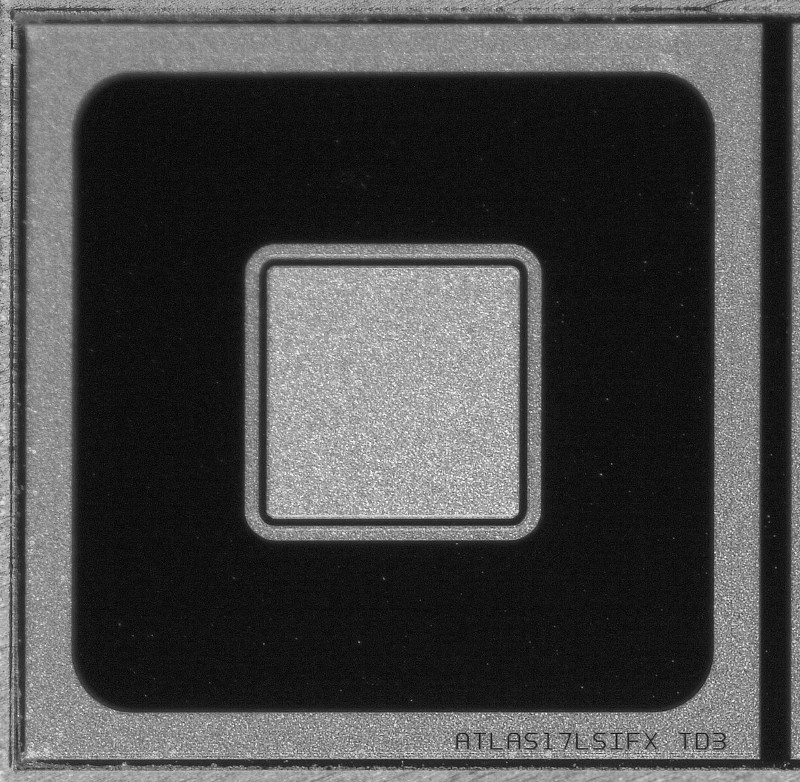}
  \caption{IFX TD3 (picture)}
  \label{fig:map2a}
\end{subfigure}
\begin{subfigure}{.23\textwidth}
  \centering
  \includegraphics[width=\linewidth]{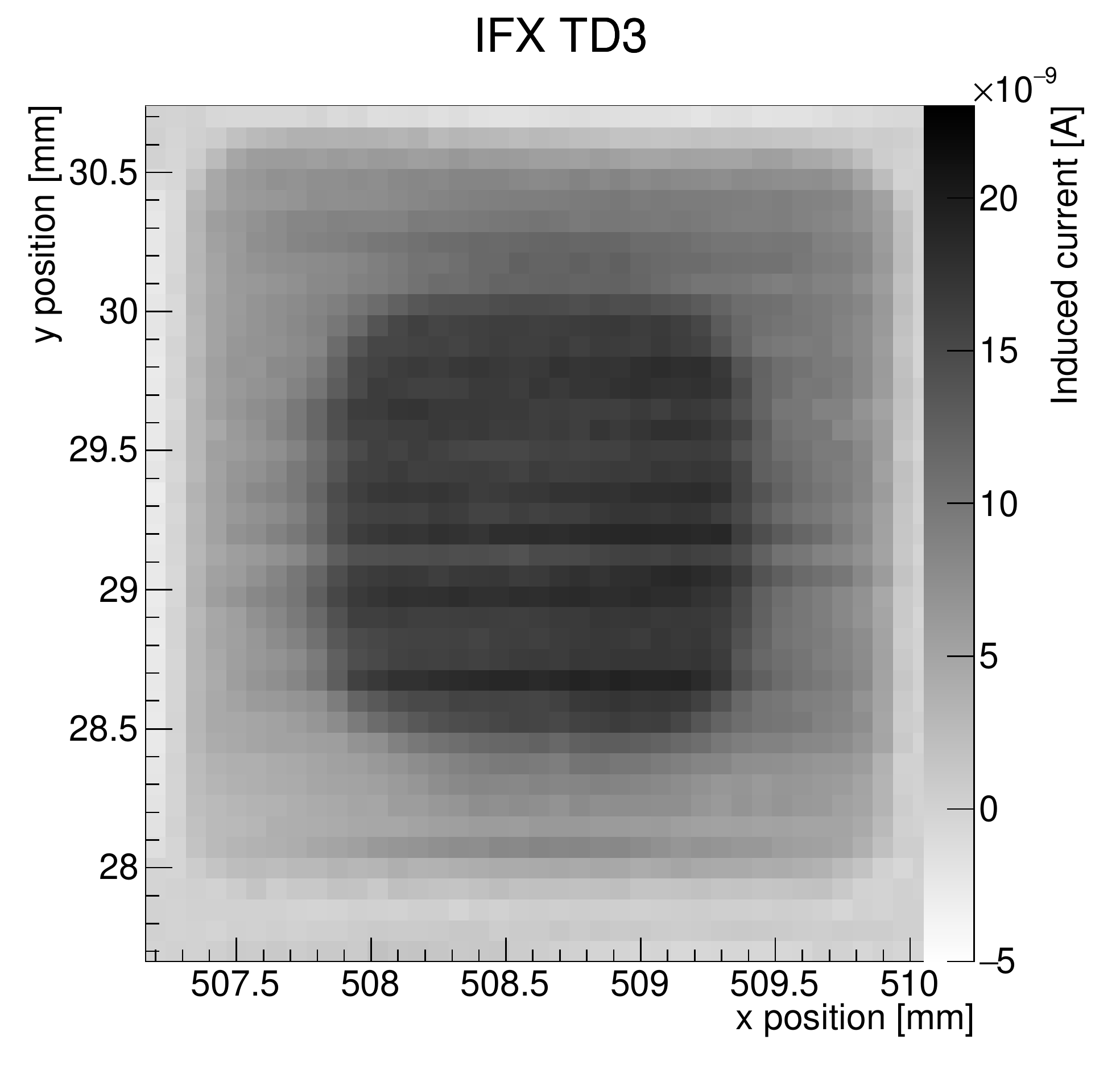}
  \caption{IFX TD3 (map)}
  \label{fig:map2b}
\end{subfigure}
\begin{subfigure}{.23\textwidth}
  \centering
  \includegraphics[width=\linewidth]{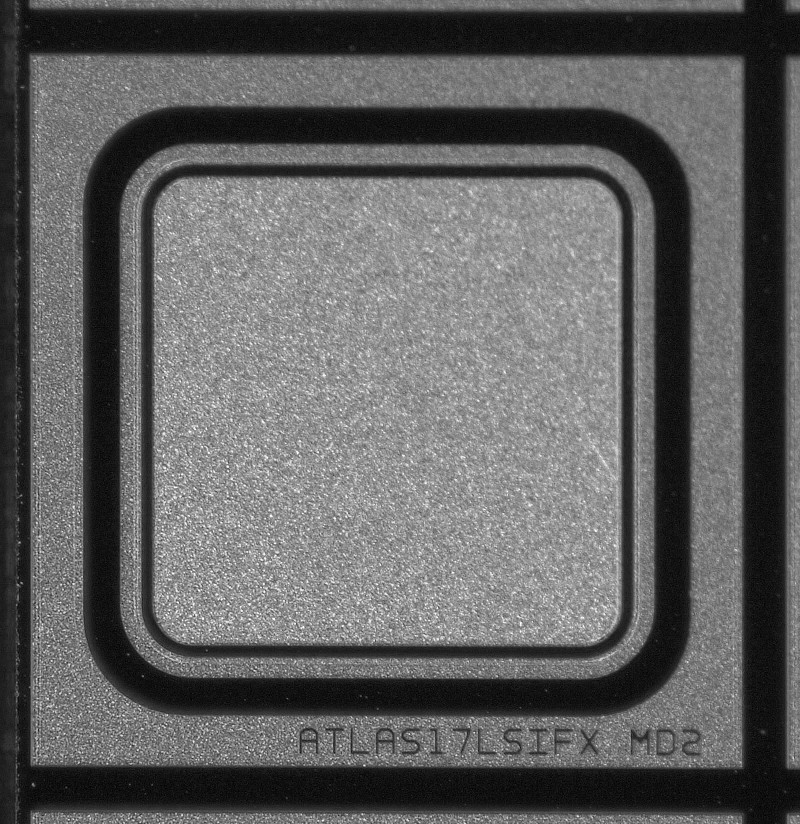}
  \caption{IFX MD2 (picture)}
  \label{fig:map3a}
\end{subfigure}
\begin{subfigure}{.23\textwidth}
  \centering
  \includegraphics[width=\linewidth]{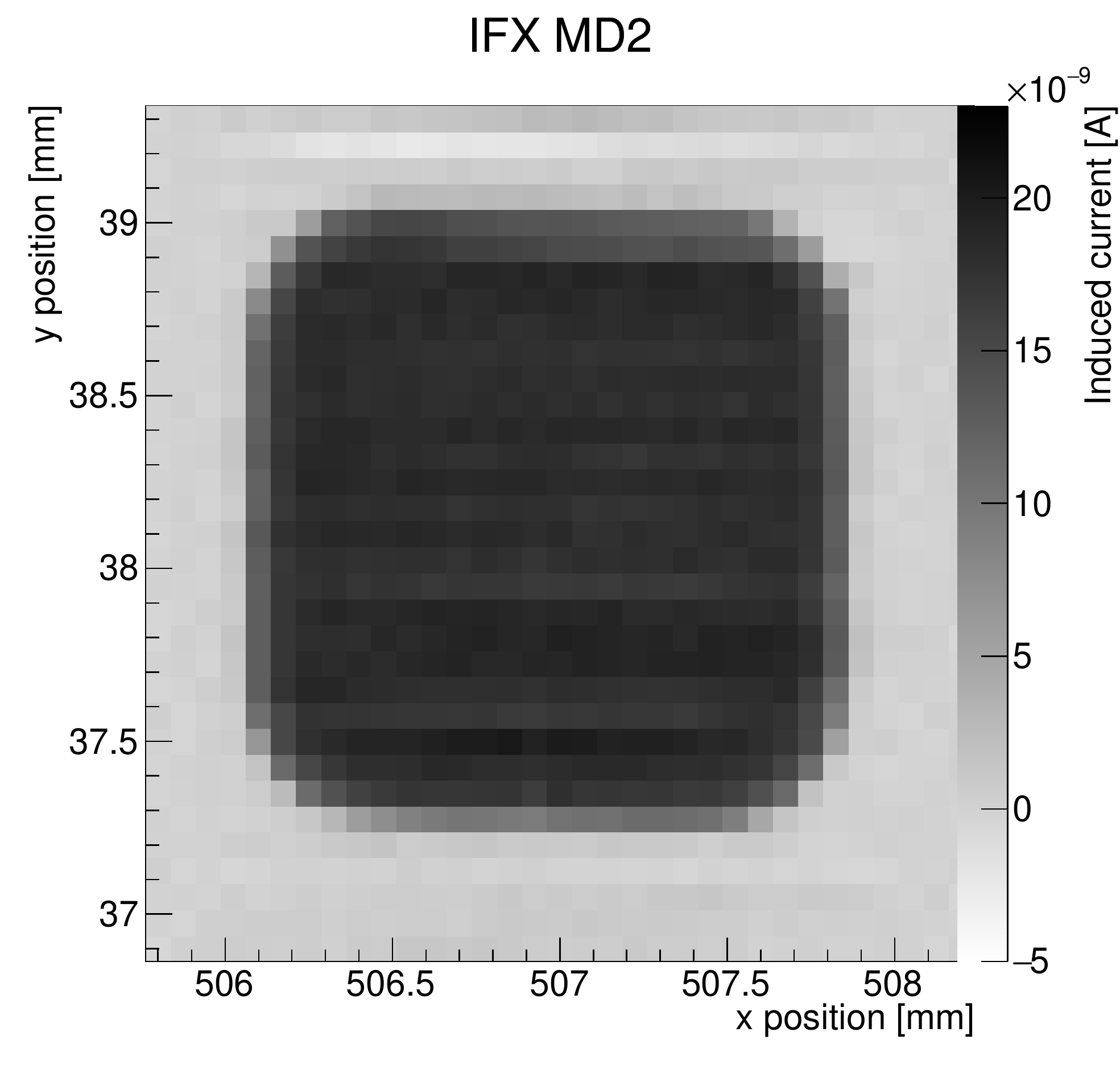}
  \caption{IFX MD2 (map)}
  \label{fig:map3b}
\end{subfigure}
\caption{Pictures and maps of diodes irradiated up to ${\unit[1\cdot10^{14}]{\text{n}_{\text{eq}}/\text{cm}^2}}$, measured at a bias voltage of {\unit[-500]{V}} each. }
\label{fig:map}
\end{figure}

Mapping the active areas of all diodes under investigation (after irradiation up to $\unit[1\cdot10^{14}]{\text{n}_{\text{eq}}/\text{cm}^2}$) showed that the shape of the depleted area matched the shape of the diode implants (see figures~\ref{fig:map1b}, \ref{fig:map2b} and~\ref{fig:map3b}). Similar as for the same diode geometries before irradiation, the shape of the active area was found to show rounded corners matching the shape of the diode edge ring (see figures~\ref{fig:map1a}, \ref{fig:map2a} and~\ref{fig:map3a}).

Samples of each diode type were irradiated up to four fluences: $\unit[1\cdot10^{14}]{\text{n}_{\text{eq}}/\text{cm}^2}$, $\unit[5\cdot10^{14}]{\text{n}_{\text{eq}}/\text{cm}^2}$, $\unit[1\cdot10^{15}]{\text{n}_{\text{eq}}/\text{cm}^2}$ and $\unit[3\cdot10^{15}]{\text{n}_{\text{eq}}/\text{cm}^2}$ (only for HPK MD2 due to time constraints). One diode per fluence was mapped in the beam, except for the highest fluence level, where the signal-to-noise-ratio was found to be too low to map the active area reliably.

\begin{figure}[htp]
\begin{subfigure}{.37\textwidth}
  \centering
  \includegraphics[width=\linewidth]{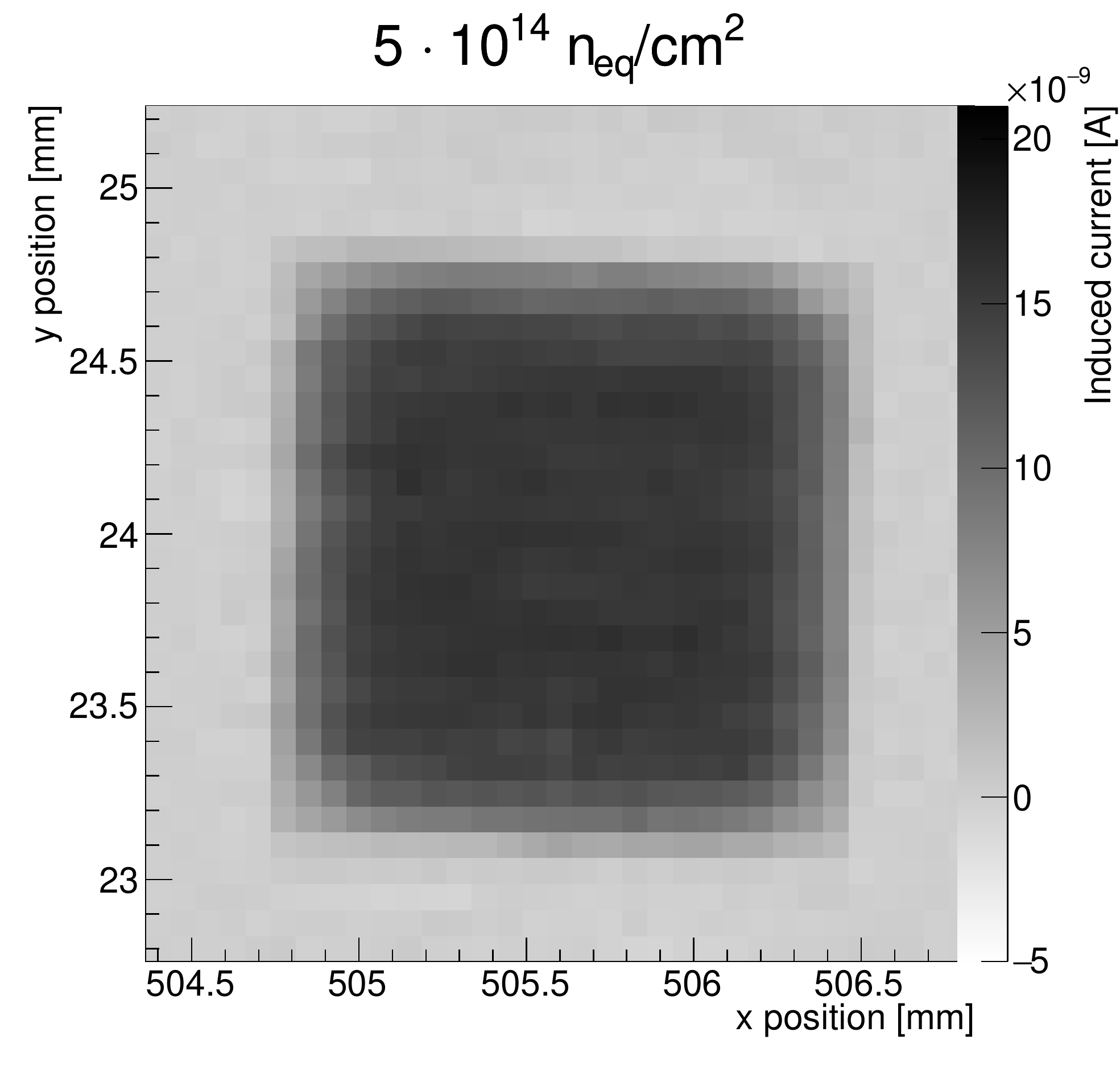}
  \caption{Map of diode HPK MD2, irradiated up to {$\unit[5\cdot10^{14}]{\text{n}_{\text{eq}}/\text{cm}^2}$}}
  \label{fig:flu2}
\end{subfigure}
\begin{subfigure}{.37\textwidth}
  \centering
  \includegraphics[width=\linewidth]{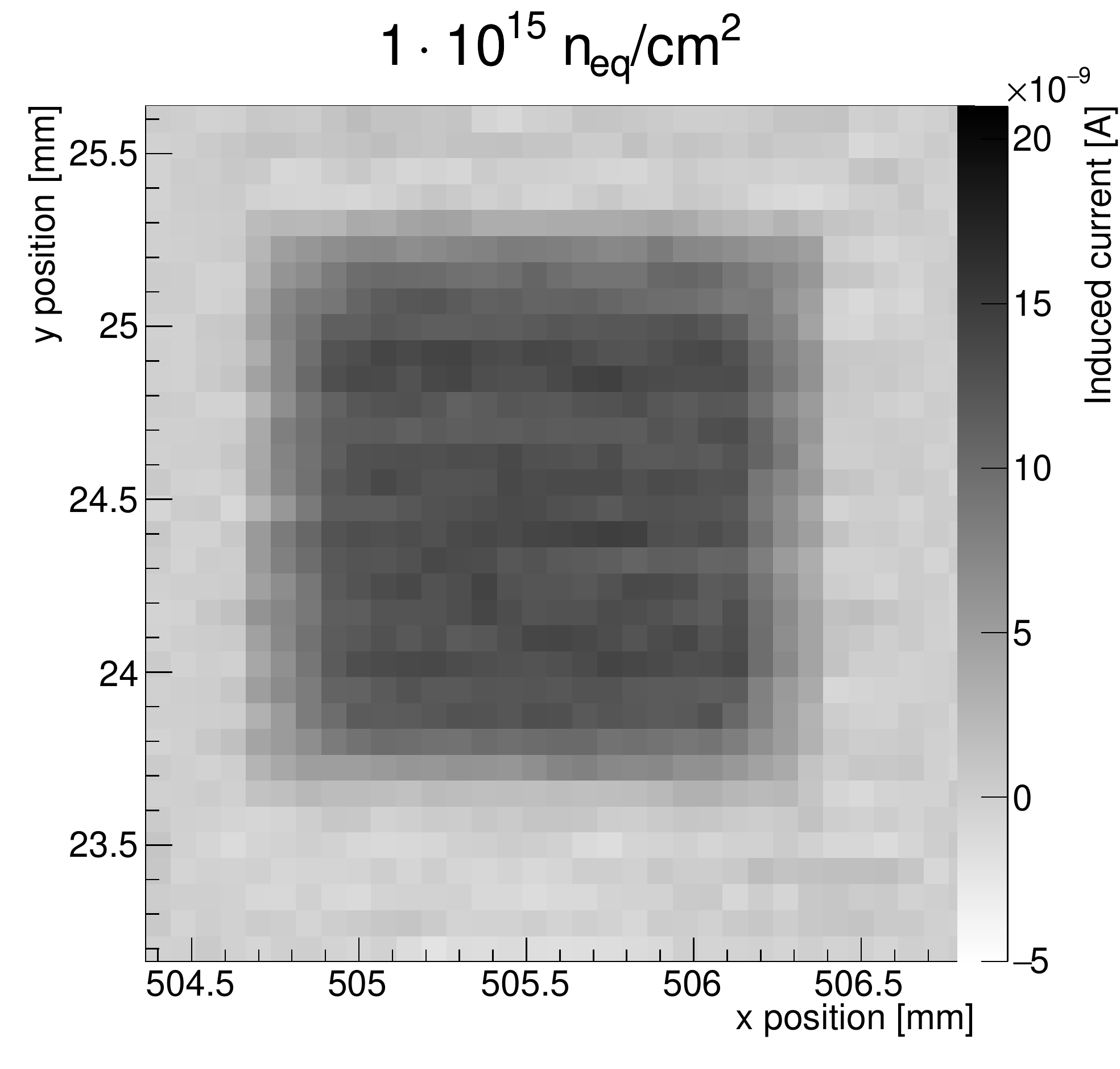}
  \caption{Map of diode HPK MD2, irradiated up to {$\unit[1\cdot10^{15}]{\text{n}_{\text{eq}}/\text{cm}^2}$}}
  \label{fig:flu3}
\end{subfigure}
\caption{Current maps for the same diode geometry, irradiated up to increasing fluences}
\label{fig:flu}
\end{figure}

The active area of diodes irradiated to different fluence levels was successfully mapped using a micro-focused X-ray beam by reading out the overall diode current.

\begin{figure}[htp]
\centering
\includegraphics[width=0.9\linewidth]{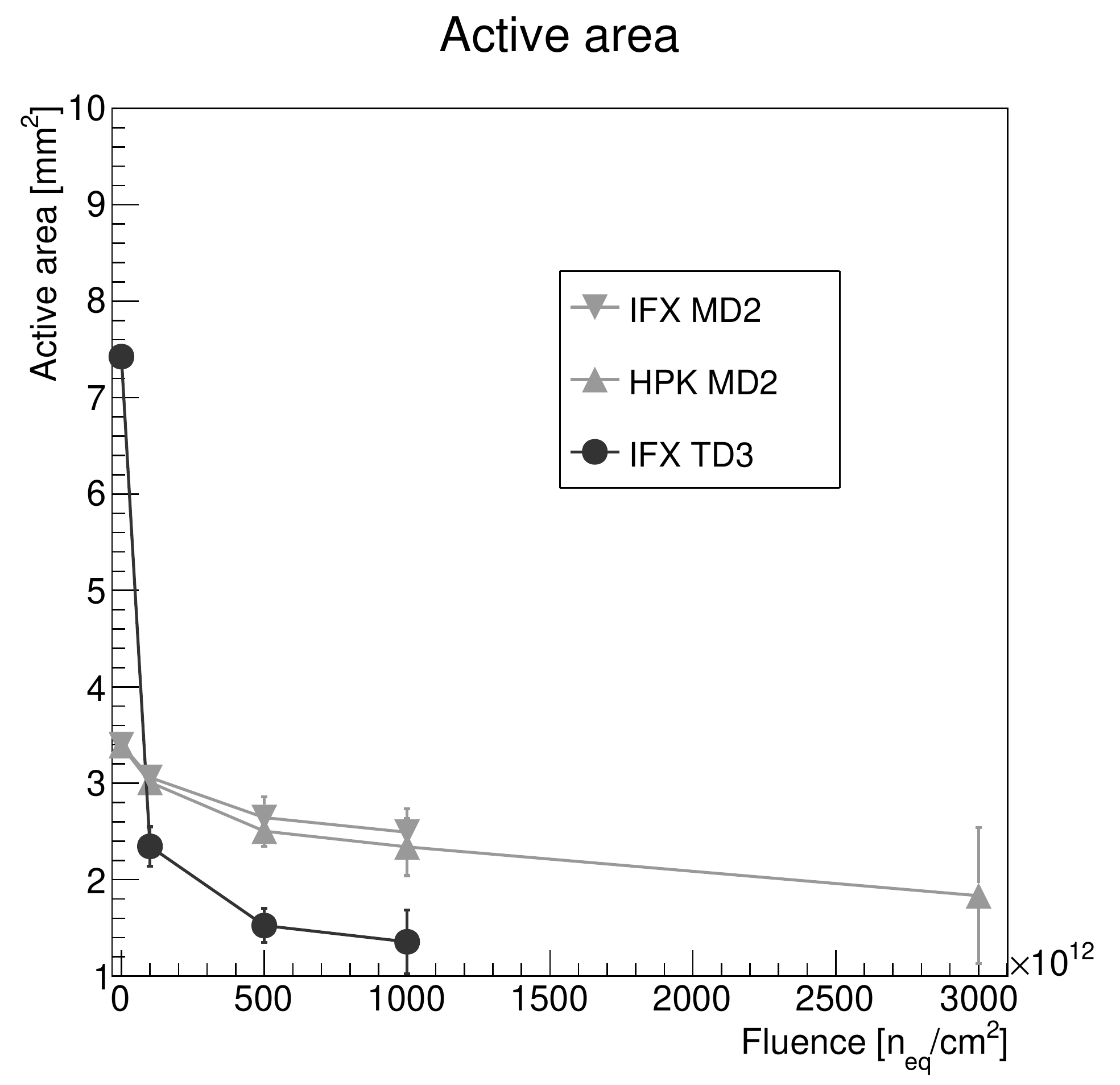}
\caption{Reduction of the active area of different diodes with increasing fluence levels. Area of unirradiated diodes from~\cite{Diodes}}
\label{fig:areas}
\end{figure}

\section{Conclusion}

The method was found suitable for measurements of the active sensor area using the total diode current. The current setup with diode temperatures of \unit[-20]{$^{\circ}$C} provided a sufficient signal-to-noise ratio to allow tests of diodes up to fluences of $\unit[1\cdot10^{15}]{\text{n}/\text{cm}^2}$ (see figures~\ref{fig:flu2} to~\ref{fig:flu3}).

\begin{table}
\scriptsize
\begin{center}
\begin{tabular}{l|ccc}
 & \multicolumn{3}{c}{Width of area, $[$mm$]$} \\
 & HPK MD2 & IFX TD3 & IFX MD2 \\
\hline
Diced size & 2 & 3 & 2\\
Bias implant & 1.14 & 1.00 & 1.25\\
\hline
Unirradiated & $1.85 \pm 0.01$ & $2.63 \pm 0.01$ & $1.96 \pm 0.01$\\
$\unit[1\cdot10^{14}]{\text{n}/\text{cm}^2}$ & $1.70 \pm 0.01$ & $1.79 \pm 0.01$ & $1.76 \pm 0.01$\\
$\unit[5\cdot10^{14}]{\text{n}/\text{cm}^2}$ & $1.55 \pm 0.01$ & $1.38 \pm 0.03$ & $1.54 \pm 0.02$\\
$\unit[1\cdot10^{15}]{\text{n}/\text{cm}^2}$ & $1.55 \pm 0.02$ & $1.22 \pm 0.01$ & $1.55 \pm 0.01$\\
$\unit[3\cdot10^{15}]{\text{n}/\text{cm}^2}$ & $1.51 \pm 0.07$ & - & -\\
\end{tabular}
\end{center}
\caption{Measured width of active diode areas for different fluences compared to the diode design layout (distance between bias ring/bias implants and overall diced diode size).}
\label{tab:sizes}
\end{table}

Measurements for all diodes showed that the active diode area shrank with increasing irradiation (see figure~\ref{fig:areas}) as expected: for an unirradiated diode, the depleted diode area extends laterally beyond the bias ring towards the dicing edge. Free charge carriers generated beyond the bias ring drift towards to the collection area and can be read out~\cite{Diodes}.
After irradiation, the diodes under investigation have decreased bulk resistivity due to an increase of acceptor-like defects, therefore the same bias voltage decreases the lateral extension of the depleted zone~\cite{explanation}. 

All diodes had depleted areas of \unit[85-95]{\%} of their diced size before irradiation, which matched the inner outline of the edge ring. Irradiation to $\unit[1\cdot10^{15}]{\text{n}/\text{cm}^2}$ shrank the active areas by \unit[31]{\%} (HPK MD2), \unit[27]{\%} (IFX MD2) and \unit[82]{\%} (IFX TD3) of the size of the unirradiated diode's active area. The performed measurements indicated that the outline of the active area before irradiation is defined by the shape of the edge ring and follows the shape and size of bias or guard ring implants after irradiation.

\section*{Acknowledgements}

This work was supported by the U.S. Department of Energy, Office of Science, High Energy Physics, under Contract No. DE-AC02-05CH11231, by USA Department of Energy, Grant DE-SC0010107, the Spanish Ministry of Economy and Competitiveness through the Particle Physics National Program, ref. FPA2015-65652-C4-4-R (MINECO/FEDER, UE), and co-financed with FEDER funds as well as by the H2020 project AIDA-2020, GA no. 654168. We acknowledge the Diamond Light Source for time on beamline B16 under proposals MT18807 and MT22002. The authors would like to thank the personnel of the B16 beam, especially Oliver Fox and Andy Malandain.

\bibliographystyle{unsrt}
\bibliography{ActiveArea.bib}

\end{document}